\documentclass[slac_one]{revtex4}
\usepackage{graphicx}
\usepackage{fancyhdr}
\begin{document}


\title{Evidence for infrared finite coupling in Sudakov resummation: a revised view-point}

\author{Georges Grunberg}
\affiliation{%
Centre de Physique Th\'eorique de l'Ecole  
Polytechnique (CNRS UMR C7644),\\
        91128 Palaiseau Cedex, France
}%


\date{\today}

\begin{abstract}

I show that Sudakov resummation takes a particularly transparent form if one deals with
the second logarithmic derivative of the short distance coefficient functions for deep inelastic scattering and 
the Drell-Yan process. A uniquely defined Sudakov exponent emerges, and I conjecture that the leftover constant
terms not included in the exponent are  given by the second logarithmic derivative of the massless quark form
factor. The meaning of a previously obtained large $N_f$ evidence for an  infrared
finite perturbative  Sudakov  coupling  is reconsidered. This coupling is  reinterpreted as a
Minkowskian coupling, making the introduction of a low-energy non-perturbative modification of the corresponding
Euclidean coupling a priori necessary. Some hints for a Banks-Zaks type of   fixed point in the Euclidean
coupling at finite $N_f$ are nevertheless pointed out, and strong evidence is provided in favor of its
universality. A criterion to select in a unique way the proper Euclidean Sudakov
 coupling relevant to the issue of power corrections is suggested. 
\end{abstract}

\maketitle

\noindent In two previous papers \cite{Gru-talk,Gru-talk1} I presented some arguments in favor of the infrared (IR)
finite coupling  approach to power
corrections in the context of Sudakov resummation. Based on new results  below, I now believe the
evidence provided in favor of an IR finite {\em perturbative Euclidean} coupling at large $N_f$ is fallacious, the
coupling having been incorrectly  identified as Euclidean. The new interpretation of the same formal results still
supports the IR finite coupling idea, but in a non-perturbative framework, closer to the spirit of the standard
\cite{DMW} approach. However some partial evidence for an IR finite  perturbative Euclidean
coupling   still exists at {\em finite}
$N_f$. The ambiguities related to the arbitrary choice of the Sudakov distribution function and effective coupling
are now resolved, and the new picture is in agreement with the IR renormalon expectation.

Consider first the scaling violation in deep inelastic scattering (DIS) in Mellin space at large $N$

\begin{equation}{d\ln F_2(Q^2,N)\over d\ln Q^2}=4 C_F\int_{0}^{Q^2}{dk^2\over k^2} G(N k^2/Q^2)
A_{{\cal S}}(k^2)+4 C_F H(Q^2)+{\cal O}(1/N)  
\label{eq:scale-viol},\end{equation}
where   the ``Sudakov effective coupling'' $A_{{\cal S}}(k^2)=a_s(k^2)+{\cal
A}_1 a_s^2(k^2)+{\cal A}_2 a_s^3(k^2)+...$, as well as $H(Q^2)$,  are  given as  power series in
$a_s\equiv \alpha_s/4\pi$ with $N$-independent coefficients. In the standard resummation framework one has
 $4 C_F A_{{\cal S}}(k^2)= A(a_s(k^2))+dB(a_s(k^2))/d\ln
k^2$, where
$A$ (the universal  ``cusp'' anomalous dimension) and $B$ are  the standard Sudakov anomalous dimensions
relevant to DIS, and
$G(N k^2/Q^2)=\exp(-N k^2/Q^2)-1\equiv G_{stan}(N k^2/Q^2)$.
Eq.(\ref{eq:scale-viol}) is equivalent to 

\begin{equation}{d\ln F_2(Q^2,N)\over d\ln Q^2}=4 C_F\left[\int_{0}^{\infty}{dk^2\over k^2} G(N k^2/Q^2)
A_{{\cal S}}(k^2)-G(\infty)\int_{Q^2}^{\infty}{dk^2\over k^2} A_{{\cal
S}}(k^2)\right]+4 C_F H(Q^2)+{\cal O}(1/N)\label{eq:ren-int-scaling-as},\end{equation}
where the two integrals on the right hand side of eq.(\ref{eq:ren-int-scaling-as}) contain only $\ln^p N$ and
constant terms, and  are free of ${\cal O}(1/N)$ terms. Since
$G(\infty)=-1$, these integrals  are separately ultraviolet (UV)
divergent, but their sum is finite. It was further  observed in \cite{Gru-talk,Gru-talk1} that the separation
between the constant terms contained in the Sudakov integral on the right hand side of eq.(\ref{eq:scale-viol})
and those contained in $H(Q^2)$ is arbitrary, different choices leading to a different ``Sudakov distribution
function'' $G(N k^2/Q^2)$ and effective coupling $A_{{\cal S}}(k^2)$, as well as to a different function
$H(Q^2)$.   The crucial new 
observation of the present paper is  that this  freedom of selecting the constant terms actually  disappears by
taking one more derivative, namely

\begin{equation}{d^2\ln F_2(Q^2,N)\over (d\ln Q^2)^2}=4 C_F\int_{0}^{\infty}{dk^2\over k^2}
\dot{G}(N k^2/Q^2) A_{{\cal S}}(k^2)+4 C_F[dH/ d\ln Q^2-A_{{\cal S}}(Q^2)]+{\cal
O}(1/N)\label{eq:d-scale-viol},\end{equation}
where $\dot{G}=- dG/ d\ln k^2$. The point is that the integral on the right hand side of
eq.(\ref{eq:d-scale-viol}) being UV convergent, all the $\ln N$ divergent terms are now determined by the
${\cal O}(N^0)$ terms contained in the integral, which therefore cannot be fixed arbitrarily anymore.
Indeed, putting 
\begin{equation}{\cal S}(Q^2,N)=\int_{0}^{\infty}{dk^2\over k^2}
\dot{G}(N k^2/Q^2) A_{{\cal S}}(k^2)\label{eq:S-exp},\end{equation}
it is easy to show that

\begin{equation}{\cal S}(Q^2,N)=c_0 a_s(Q^2)+(\beta_0 c_0 L + {\cal A}_1 c_0-\beta_0 c_1)
a_s^2(Q^2)+[\beta_0^2 c_0 L^2 +(2\beta_0 ({\cal A}_1 c_0-\beta_0 c_1)+\beta_1 c_0)L+{\cal O}(L^0)]a_s^3(Q^2)+...
\label{eq:log-structure},\end{equation}
where $L=\ln N$ and $c_p=\int_{0}^{\infty}{d\epsilon\over \epsilon} \dot{G}(\epsilon)\ln^p(\epsilon)$. Thus $c_0$
determines all the leading logs of $N$ (which implies that $c_0=1$), while the combination  ${\cal A}_1 c_0-\beta_0
c_1$ determines the sub-leading logs and is therefore fixed, etc...Although the ``Sudakov exponent'' ${\cal
S}(Q^2,N)$ is now uniquely determined, the Sudakov distribution function and effective coupling are not. For
instance the value of ${\cal A}_1$ is correlated to that of $c_1$. In fact, one can look at eq.(\ref{eq:S-exp})
for any given choice of the Sudakov
distribution function
$\dot{G}(N k^2/Q^2)$  as defining an  integral transform mapping the Sudakov effective coupling $A_{{\cal
S}}(k^2)$ to the Sudakov exponent ${\cal S}(Q^2,N)$. The only constraint on the transform is the normalization
$c_0=1$. From this point of view, all choices of $\dot{G}$ are equivalent, and the very question \cite{Gru-talk1}
whether
$A_{{\cal S}}(k^2)$ should be identified to an Euclidean or to a Minkowskian coupling appears meaningless at this
stage. Only additional physical information (to be provided below) can help clarify this point.
 
Moreover, since the Sudakov exponent ${\cal S}(Q^2,N)$ is  uniquely determined, so is the combination
$dH/ d\ln Q^2-A_{{\cal S}}(Q^2)$ of the ``leftover'' constant terms not included in the Sudakov exponent. In
fact, I conjecture that it is related to the space-like on-shell electromagnetic quark form factor
\cite{Magnea-Sterman,MVV} ${\cal F}(Q^2)$ by
  
\begin{equation}4 C_F\left({dH\over d\ln Q^2}-A_{{\cal S}}(Q^2)\right)={d^2\ln \vert{\cal F}(Q^2)\vert^2 \over
(d\ln Q^2)^2}\label{eq:form-factor}.\end{equation}
The fact that the second logarithmic derivative of the form factor is both finite and renormalization group
invariant follows from the properties \cite{Collins} of the evolution equation   satisfied by the form factor. I
have checked that eq.(\ref{eq:form-factor}) is satisfied to ${\cal O}(a_s^3)$ in the large $N_f$ limit. Further
checks to the same order at finite $N_f$, as well as to all orders at large $N_f$, are under consideration.

For the short distance Drell-Yan cross section, the analogues of eq.(\ref{eq:d-scale-viol}) and
(\ref{eq:form-factor}) are

\begin{equation}{d^2\ln \sigma_{DY}(Q^2,N)\over (d\ln Q^2)^2}=4 C_F\int_{0}^{\infty}{dk^2\over k^2}
\dot{G}_{DY}(N k/Q) A_{{\cal S},DY}(k^2)+4 C_F[dH_{DY}/ d\ln Q^2-A_{{\cal S},DY}(Q^2)]+{\cal
O}(1/N)\label{eq:d-scale-viol-DY},\end{equation}
where $\dot{G}_{DY}=- dG_{DY}/ d\ln k^2$, with $G_{DY}(N k/Q)=\exp(-N k/Q)-1\equiv G_{DY}^{stan}(N k/Q)$
and $4 C_F A_{{\cal S},DY}(k^2)=  A(a_s(k^2))+{1\over 2}{dD(a_s(k^2))\over d\ln
k^2}$ within the standard resummation framework ($D$ is the usual D-term), 
and

\begin{equation}4 C_F\left({dH_{DY}\over d\ln Q^2}-A_{{\cal S},DY}(Q^2)\right)={d^2\ln \vert{\cal F}(-Q^2)\vert^2
\over (d\ln Q^2)^2}\label{eq:form-factor-DY},\end{equation}
where ${\cal F}(-Q^2)$ is the time-like quark form factor. Eq.(\ref{eq:form-factor}) and (\ref{eq:form-factor-DY})
are in the same vein as  similar results \cite{Sterman, Magnea, L-M} relating the ratio of
time-like to space-like form factors to the Drell-Yan cross-section normalized to the (square) of the DIS one.

I now come to the question of the physical interpretation of the Sudakov effective coupling $A_{{\cal S}}(k^2)$,
and the way to eliminate the non-uniqueness in its definition: these issues are crucial for a proper understanding
of the  power corrections arising within the IR finite coupling approach. One would like to associate $A_{{\cal
S}}(k^2)$ to some kind of dressed gluon propagator, which does not seem possible in general, given that the
resummation formulas eq.(\ref{eq:d-scale-viol}) or (\ref{eq:d-scale-viol-DY})  are valid beyond the single gluon
exchange approximation. However, there is a definite limit in QCD, the large $N_f$ limit, where the latter
approximation naturally arises. In particular,  the dispersive approach \cite{DMW,BB}, which usually provides
the most convenient calculational technique in this limit, allows to identify a physical {\em Minkowskian}
coupling, namely the time-like (integrated) discontinuity of the {\em Euclidean} one-loop coupling (the so-called
``V-scheme'' coupling) associated to the  dressed gluon propagator, given by

\begin{equation}A_{{\cal
S}}^{Mink}(k^2)={1\over\beta_0}\left[{1\over
2}-{1\over\pi}\arctan(t/\pi)\right]\label{eq:A-simple},\end{equation}
with $t=\ln (k^2/\Lambda^2_V)$ (where $\Lambda_V$ is the V-scheme scale parameter). It is then natural to select
the Sudakov distribution function by the requirement that the associated Sudakov effective coupling is just given
by
$A_{{\cal S}}^{Mink}(k^2)$ at {\em large} $N_f$. As shown in \cite{Gru-talk,Gru-talk1}, this requirement fixes the
corresponding ``Minkowskian''  Sudakov distribution function (which one could also call ``characteristic
function'' following
\cite{DMW}) to be given in the DIS case by
$G_{Mink}(N k^2/Q^2)=
\ddot{{\cal G}}_{SDG}(\epsilon)$, with
$\epsilon=N k^2/Q^2$ and
\begin{equation} \ddot{{\cal G}}_{SDG}(\epsilon)=G_{stan}(\epsilon)-{1\over 2} \epsilon\
\exp(-\epsilon)-{1\over 2} \epsilon\
\Gamma(0,\epsilon)+{1\over 2} \epsilon^2\
\Gamma(0,\epsilon)
\label{eq:Gdot-SDG},\end{equation}
where $\Gamma(0,\epsilon)$ is the incomplete gamma function. In the Drell-Yan case, the same requirement yields 
instead

\begin{equation}G_{Mink}^{DY}(N k/Q)=2 {d\over d\ln Q^2}\left[K_0(2
N k/Q)+\ln(N k/Q)+\gamma_E\right]=\ddot{{\cal G}}_{SDG,DY}(\epsilon_{DY}^2)\label{eq:G-DY-new},\end{equation}
where $K_0$ is the modified Bessel function of the second kind, and $\epsilon_{DY}=N k/Q$. This result also
follows from the resummation formalism (not tied to the single gluon approximation) of
\cite{Vogelsang}, which  therefore uses an implicitly Minkowskian framework in the above sense.
 
However, although
the  coupling $A_{{\cal S}}^{Mink}(k^2)$ of eq.(\ref{eq:A-simple}) is IR finite at the perturbative level, it
{\em cannot} be taken as an evidence in favor of the IR finite coupling approach to power corrections, contrary to
the statements in \cite{Gru-talk, Gru-talk1}, since it is now clear it should be identified to a Minkowskian
coupling. The corresponding Euclidean coupling $A_{{\cal S}}^{Eucl}(k^2)$ at large $N_f$  is just the one-loop
V-scheme coupling 
\begin{equation}A_{{\cal S}}^{Eucl}(k^2)={1\over \beta_0 \ln(k^2/\Lambda_V^2)}\label{eq:one-loop},\end{equation} 
which has a Landau pole, and
thus by itself provides no  evidence in favor of the IR finite coupling approach (which relies in an essential way
\cite{DMW,Gru-power} on the IR finitness of the {\em Euclidean} coupling). The {\em assumption} of IR
finitness must therefore be made, as usual \cite{DMW}, at the {\em non-perturbative} level, by postulating the
existence of a non-perturbative modification $\delta A_{{\cal S}}^{Eucl}(k^2)$ of the Euclidean coupling at low
scales. There is nevertheless some (admitedly not yet conclusive) indication for the existence of an IR fixed point
in the perturbative Euclidean coupling at {\em finite} $N_f$. Indeed the three-loop Sudakov effective coupling
beta-function $dA_{{\cal S}}^{Eucl}/d\ln k^2=-\beta_0 (A_{{\cal S}}^{Eucl})^2-\beta_1 (A_{{\cal
S}}^{Eucl})^3-\beta_2^{Eucl} (A_{{\cal S}}^{Eucl})^4+...$
 does have an IR fixed point even at low $N_f$ values, due to a large negative three-loop coefficient
$\beta_2^{Eucl}$. For instance at $N_f=4$ one gets $4\pi A_{{\cal S},IR}^{Eucl}\simeq 0.6 $ in the DIS case, which
looks marginally perturbative (a similar value is obtained in the Drell-Yan case). Of course, as in other examples
\cite{G-K}, this fixed point could be easily washed out by 4-loop corrections. Indeed, although the standard
Banks-Zaks expansion
\cite{G-K,Gru-condensate} of $A_{{\cal S},IR}^{Eucl}$ in powers of $16.5-N_f$   appears divergent at  $N_f=4$, the
modified expansion suggested in
\cite{Gru-window} yields a reasonably small next-to-leading order correction. The issue of an IR finite
perturbative Euclidean Sudakov coupling is thus still open.

\noindent The IR power corrections are best parametrized
\cite{Gru-power,Euclidean}  in term of low energy moments of the  Euclidean coupling, and it is therefore  useful
to introduce the corresponding Euclidean Sudakov distribution function $G_{Eucl}$. In the DIS case, the latter is
related  to the Minkowskian  distribution function by the dispersion relation

\begin{equation} G_{Mink}(\epsilon)=\epsilon\int_0^{\infty} dy {G_{Eucl}(y)\over (\epsilon+y)^2}
\label{eq:disp}.\end{equation}
Actually, in the DIS case, the Euclidean Sudakov distribution turns out not to exist.
This fact is related to the circumstance that the discontinuity of the corresponding Sudakov characteristic
function (eq.(\ref{eq:Gdot-SDG})) for any finite $\epsilon<0$  is of the form
$a \epsilon^2+b \epsilon^4$ (yielding only two \cite{Gardi-Roberts} power corrections in the Sudakov exponent).
One   then has to rely on the procedure of \cite{Gru-power,Euclidean} to relate IR power corrections to moments of
the Euclidean coupling.

\noindent In the Drell-Yan case,
one gets 
\begin{equation}G_{Eucl}^{DY}(\epsilon_{DY})=J_0(2\epsilon_{DY})-1\equiv
\tilde{G}_{Eucl}^{DY}(\epsilon_{DY}^2)\label{eq:G-DY-Eucl},\end{equation}  where $J_0$ is the Bessel function, an
{\em even} function of $\epsilon_{DY}=N k/Q$: this property ensures only even power corrections (with no
logarithmic enhancement) are present, in agreement with the renormalon argument
\cite{Beneke-Braun}. The analogue of eq.(\ref{eq:disp}) is

\begin{equation} \tilde{G}_{Mink}^{DY}(\epsilon_{DY}^2)=\epsilon_{DY}^2\int_0^{\infty} dy
{\tilde{G}_{Eucl}^{DY}(y)\over (\epsilon_{DY}^2+y)^2}
\label{eq:disp-DY}.\end{equation}
where (eq.(\ref{eq:G-DY-new})) $\tilde{G}_{Mink}^{DY}(\epsilon_{DY}^2)\equiv\ddot{{\cal
G}}_{SDG,DY}(\epsilon_{DY}^2)$.

\noindent The Euclidean or Minkowskian Sudakov distribution functions fixed through the large $N_f$ identification
of the Sudakov effective couplings can then be used to determine the corresponding  effective couplings at {\em
finite}
$N_f$ in the usual way, requiring  the divergent $\ln^p N$ terms to be correctly reproduced order by order in
$\alpha_s$. It is  natural to keep referring to the resulting
$A_{{\cal S}}^{Eucl}(k^2)$ and $A_{{\cal S},DY}^{Eucl}(k^2)$ couplings (or $A_{{\cal S}}^{Mink}(k^2)$ and
$A_{{\cal S},DY}^{Mink}(k^2)$ couplings)
  as Euclidean (resp. Minkowskian) couplings even at {\em finite} $N_f$, where
 identification to a dressed gluon propagator is no longer possible. 

\noindent\underline{Universality issues}: I note that at large $N_f$ there is
universality to all orders in $\alpha_s$ between $A_{{\cal S}}^{Eucl}(k^2)$ and $A_{{\cal S},DY}^{Eucl}(k^2)$,
since they are both equal to the  one-loop V-scheme coupling in this limit. At finite $N_f$ however it easy to
check that  universality in the ultraviolet region  holds only up to next to leading order, where the Euclidean 
Sudakov effective couplings actually coincide with the ``cusp'' anomalous dimension \cite{footnote1}, but is lost
beyond that order.

\noindent On the other hand, an interesting universality property holds in the IR region at large enough, but
finite
$N_f$. Indeed for $N_f$ close  to the value $16.5$ where asymptotic freedom is lost, there is (as already
mentioned) a Banks-Zaks type of  fixed point, which is expected to persist within the so-called ``conformal
window'', whose lower boundary might eventually extend down to $N_f$ values as low \cite{Gru-fixing} as $N_f=4$.
I found that the Banks-Zaks expansions of the IR fixed point are {\em identical} for the Sudakov effective
couplings relevant to DIS and Drell-Yan up to the
$N^3LO$ order! Moreover, this property is independent of the ambiguity in choice of the Sudakov effective
couplings. This finding  suggests that the Banks-Zaks fixed point might be identical for the DIS and
Drell-Yan Sudakov effective couplings to all orders, which strongly supports the (approximate) universality of the
corresponding IR power corrections.

\acknowledgments
I wish to thank Yuri Dokshitzer, Mrinal Dasgupta and Gavin Salam for having organized this very stimulating
 workshop. I am indebted to S. Friot for the result quoted in  eq.(\ref{eq:G-DY-Eucl}).

\bibliography{apssamp}

\begin{thebibliography}{9}
\bibitem{footnote1} This is a non-trivial statement as far as the DIS
coupling $A_{{\cal S}}^{Eucl}(k^2)$ is concerned. In particular, it is not satisfied by the coupling $A_{{\cal
S}}(k^2)$ resulting from the standard resummation formalism.
\end{thebibliography}

\end{document}